\documentclass[11pt]{article}
\usepackage[margin=1in]{geometry}
\usepackage{graphicx}
\usepackage{amsmath}
\usepackage{amssymb}
\usepackage{booktabs}
\usepackage{multirow}
\usepackage{hyperref}
\usepackage{cite}
\usepackage{xcolor}

\title{Spectrogram Patch Codec: A 2D Block-Quantized VQ-VAE and HiFi-GAN for Neural Speech Coding}
\author{
    \begin{minipage}[t]{0.45\textwidth}
      \centering
      Luís Felipe Chary \\
      \textit{Department of Electronic Systems Enginnering} \\
      \textit{Universidade de São Paulo} \\
      \texttt{luisfchary@usp.br}
    \end{minipage}
    \and
    \begin{minipage}[t]{0.45\textwidth}
      \centering
      Miguel Arjona Ramirez \\
      \textit{Department of Electronic Systems Enginnering} \\
      \textit{Universidade de São Paulo} \\
      \texttt{maramire@usp.br}
    \end{minipage}
}
\date{}

\begin{document}

\maketitle

\begin{abstract}
We present a neural speech codec that challenges the need for complex residual vector quantization (RVQ) stacks by introducing a simpler, single-stage quantization approach. Our method operates directly on the mel-spectrogram, treating it as a 2D data and quantizing non-overlapping 4×4 patches into a single, shared codebook. This patchwise design simplifies the architecture, enables low-latency streaming, and yields a discrete latent grid. To ensure high-fidelity synthesis, we employ a late-stage adversarial fine-tuning for the VQ-VAE and train a HiFi-GAN vocoder from scratch on the codec's reconstructed spectrograms. Operating at approximately 7.5 kbits/s for 16 kHz speech, our system was evaluated against several state-of-the-art neural codecs using objective metrics such as STOI, PESQ, MCD, and ViSQOL. The results demonstrate that our simplified, non-residual architecture achieves competitive perceptual quality and intelligibility, validating it as an effective and open foundation for future low-latency codec designs.
\end{abstract}

\section{Introduction}
Neural speech codecs compress signals into compact, perceptually meaningful discrete tokens that can be streamed with low latency. Vector‑quantized autoencoders (VQ‑VAEs) discretize a learned representation, alleviating posterior collapse and exposing an \emph{acoustic language} of codes~\cite{oord2017vqvae}. Recent systems such as SoundStream~\cite{zeghidour2021soundstream} and EnCodec~\cite{defossez2022encodec} use residual vector quantization (RVQ) and adversarial training to reach 3--24~kbits/s with strong quality. However, RVQ stacks can complicate deployment and memory use.

We propose a codec that \emph{does not} relies on residual quantization: instead, it operates on the mel‑spectrogram as 2D information and quantizes $4\times4$ patches into a single shared codebook. This yields a simple, low‑latency 2D grid of tokens with a controllable bitrate (Sec.~\ref{sec:bitrate}). We adopt a late adversarial stage and train a HiFi‑GAN \cite{kong2020hifigan} vocoder \emph{from scratch} on the codec's reconstructions, improving the match between codec artifacts and the vocoder's generation. Our work is also informed by recent evidence that strong autoregressive decoders remain highly competitive with diffusion under comparable budgets~\cite{sun2024autoregressivemodelbeatsdiffusion}.

\paragraph{Contributions.}
\begin{itemize}
  \item A \textbf{single‑level, 2D block‑quantized} VQ‑VAE speech codec (no RVQ), quantizing non‑overlapping $4\times4$ mel patches into a $K{=}4096$ codebook, producing a $(T/4)\times(80/4)$ grid of discrete tokens.
  \item \textbf{Late adversarial fine‑tuning} for the VQ-VAE and a \textbf{HiFi‑GAN vocoder trained from scratch} on reconstructed spectrograms from the codec, improving perceptual realism with stable training.
  \item A \textbf{bitrate derivation} matching our configuration (Sec.~\ref{sec:bitrate}), and \textbf{fair comparisons} to DAC~\cite{kumar2023dac}, EnCodec~\cite{defossez2022encodec}, and SNAC~\cite{siuzdak2024snac} at nearby rates.
  \item \textbf{Objective evaluations} using STOI~\cite{taal2011stoi}, MCD~\cite{kubichek1993mcd}, PESQ~\cite{rix2001pesq}, ViSQOL~\cite{chinen2020visqol}, and RTF on a held‑out test set, highlighting the trade‑off between bitrate, intelligibility and perceptual quality.
  \item \textbf{Design for 16~kHz sampled speech}: the entire system operates on 16~KHz sampled audio with 80‑band mel spectrograms, making it suitable for low‑latency speech applications.
\end{itemize}

\section{Related Work}
\paragraph{VQ‑VAEs and neural codecs.}
VQ‑VAE~\cite{oord2017vqvae} discretizes latent spaces via vector quantization, widely used for image, audio, and video. Neural audio codecs based on RVQ include SoundStream~\cite{zeghidour2021soundstream} and EnCodec~\cite{defossez2022encodec}, reporting 3--24~kbits/s in real time. The Descript Audio Codec (DAC)~\cite{kumar2023dac} introduces improved RVQGAN techniques and achieves high‑fidelity compression of 44.1~kHz sampled audio at 8~kbits/s. SNAC~\cite{siuzdak2024snac} proposes a multi‑scale neural audio codec built on RVQ that employs quantizers at different temporal resolutions.

\paragraph{Autoregressive vs.~diffusion.}
Recent analyses indicate that carefully tuned autoregressive models can \emph{beat} or match diffusion in several regimes at comparable compute~\cite{sun2024autoregressivemodelbeatsdiffusion}, reinforcing the relevance of token‑based codecs paired with AR decoders for audio processing models.

\paragraph{Vocoder models.}
HiFi‑GAN~\cite{kong2020hifigan} is a fast adversarial vocoder from mel‑spectrograms, widely adopted for low‑latency synthesis. In contrast to prior work, we \emph{train HiFi‑GAN from scratch} use spectrograms \emph{reconstructed by our codec}, rather than clean representations, to better match codec artefacts. The ViSQOL family of metrics provides an open‑source, full‑reference objective assessment of speech quality; the v3 release emphasises that PESQ and POLQA are standards for speech quality and positions ViSQOL as a freely available alternative that is continually extended to new domains~\cite{chinen2020visqol}.

\section{Codec Architecture}
Given a 16~kHz sampled waveform, we compute an $F{=}80$‑band mel‑spectrogram $x\in\mathbb{R}^{T\times F}$ with a short‑time Fourier transform (STFT) window of 512 and hop of 128. The VQ-VAE consists of an encoder, a vector codebook, and a decoder. The encoder is a convolutional network that downsamples the representation by $4\times$ and outputs a latent feature map. A $4\times4$ patchification layer tiles $x$ into non‑overlapping patches along $(t,f)$. Each patch is projected and \emph{directly} quantized by a single codebook of size $K{=}4096$ (12~bits/index). The resulting discrete 2D grid has shape $(T/4)\times(F/4)=(T/4)\times20$. The decoder mirrors the encoder to reconstruct $\hat{x}$.

Finally, a HiFi-GAN vocoder synthesizes the final waveform from the reconstructed spectrograms. The HiFi-GAN generator was adapted for 16~KHz sampled audio, using four blocks of transposed convolutions for upsampling with kernel sizes of 16, 16, 4, 4 and corresponding strides of 4, 4, 4, 2. This configuration results in a total upsampling factor of 128, matching our STFT hop size. The discriminators, including the Multi-Scale Discriminator (MSD) and Multi-Period Discriminator (MPD), follow the exact design and hyperparameters from the original HiFi-GAN paper~\cite{kong2020hifigan}.

\begin{figure}[!ht]
\centering
\includegraphics[width=.9\textwidth]{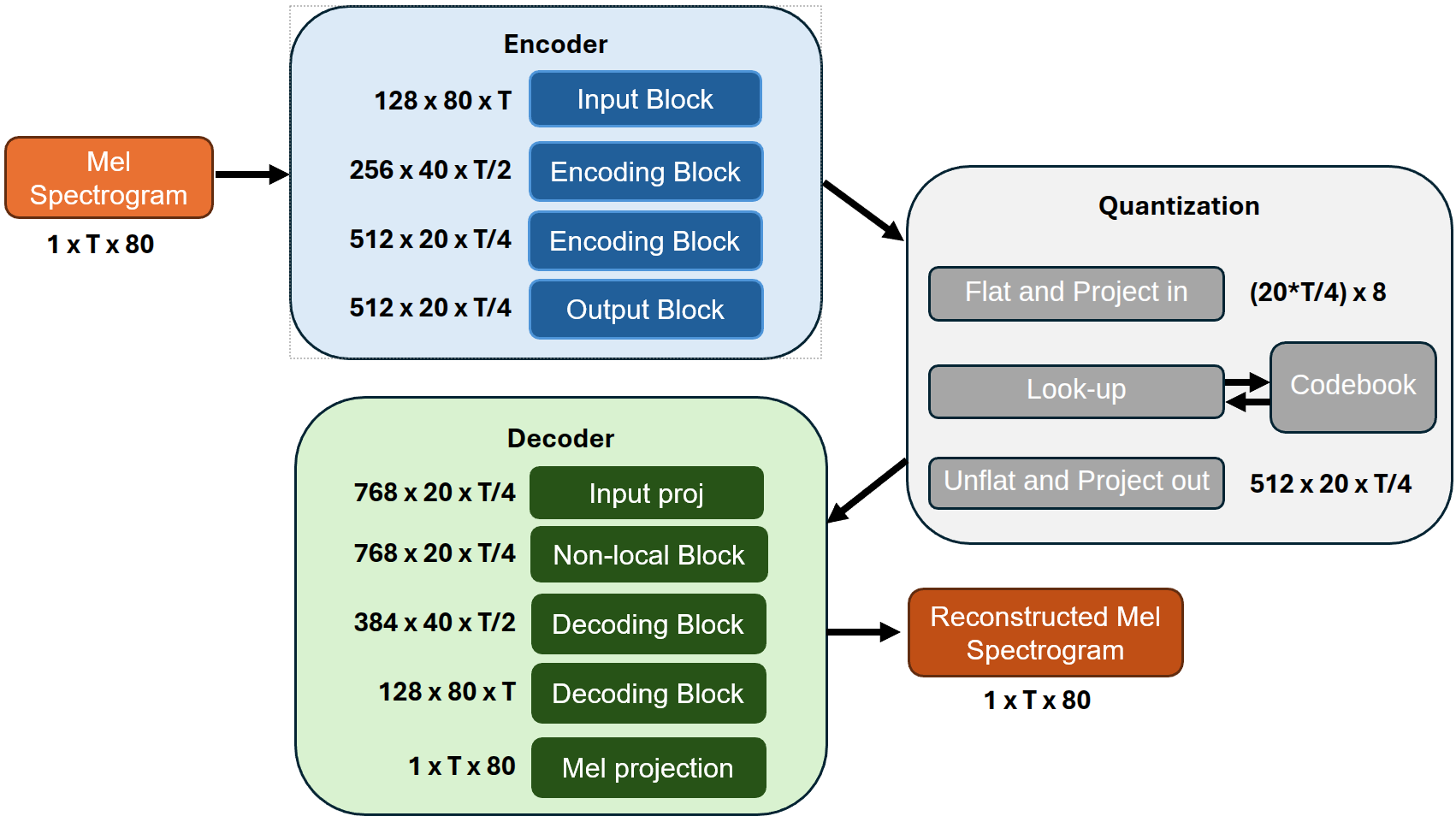}
\caption{Data flow of the proposed VQ-VAE architecture. The Encoder progressively downsamples the input mel spectrogram. The Quantization block reshapes the latent tensor, performs a codebook look-up for discretization, and prepares the result for the Decoder, which symmetrically reconstructs the spectrogram. Tensor dimensions are indicated throughout the pipeline.}
\label{fig:arch}
\end{figure}

\subsection{quantization and losses}
\label{sec:quantization}
Let $x$ be partitioned into non‑overlapping patches $p_{i,j}\in\mathbb{R}^{4\times4}$ with $(i,j)\in\{1..T/4\}\times\{1..20\}$. Each patch is embedded and replaced by its nearest codebook vector $e_{k_{i,j}}$, $k_{i,j}\in\{1..K\}$. Denote by $z_{i,j}$ the encoder output before quantization for patch $(i,j)$ and by $\mathrm{sg}(\cdot)$ the stop‑gradient operator. Our codec is trained with three loss components:

\paragraph{Reconstruction loss.} The mel‑spectrogram reconstruction loss combines an $\ell_1$ term and a perceptual LPIPS \cite{zhang2018perceptual} term:
\begin{equation}
\mathcal{L}_{\mathrm{rec}} = \|x - \hat{x}\|_1 + \mathrm{LPIPS}(x,\hat{x}).
\end{equation}

\paragraph{Vector‑quantization loss.} To train the codebook, we follow the original VQ‑VAE formulation~\cite{oord2017vqvae}. For each latent patch $(i,j)$ we define
\begin{equation}
\mathcal{L}_{\mathrm{VQ}} = \sum_{i,j} \left\| \mathrm{sg}(z_{i,j}) - e_{k_{i,j}} \right\|_2^2 + \beta \, \left\| z_{i,j} - \mathrm{sg}(e_{k_{i,j}}) \right\|_2^2,
\end{equation}
where the first term updates the codebook embeddings and the second ("commitment") term encourages the encoder to commit to the selected codebook entries. We set $\beta=0.25$ in our experiments.

\paragraph{Adversarial loss.} We introduce a PatchGAN-style \cite{isola2017pix2pix} spectrogram discriminator $D$ for the VQ-VAE. Let $G$ be the generator (encoder-decoder). The adversarial loss for the generator is:
\begin{equation}
\mathcal{L}_{\mathrm{GAN}} = \lambda_{\mathrm{adv}} \, D(G(x)),
\end{equation}
where $\lambda_{\mathrm{adv}}$ is adjusted dynamically based on the ratio of reconstruction to GAN gradient norms to ensure stable training. The total loss during adversarial fine-tuning is:
\begin{equation}
\mathcal{L}_{\mathrm{total}} = \mathcal{L}_{\mathrm{rec}} + \mathcal{L}_{\mathrm{VQ}} + \mathcal{L}_{\mathrm{GAN}}.
\end{equation}

\section{Training Procedure}
We use a dedicated multi-stage training procedure for the VQ-VAE and the HiFi-GAN vocoder.

\subsection{VQ-VAE Training}
The VQ-VAE is trained for 150,000 steps using the AdamW \cite{loshchilov2019adamw} optimizer with an initial learning rate of $3\times10^{-4}$, $\beta_1=0.9$, $\beta_2=0.95$, and a weight decay of 0.01. The batch size is set to 256. We use a learning rate schedule with a linear warm-up for the first 1,000 steps, followed by a cosine decay \cite{loshchilov2017sgdr}. To stabilize training, the adversarial discriminator and its corresponding loss are only introduced after 20,000 steps.

\subsection{HiFi-GAN Training}
The HiFi-GAN vocoder is trained separately \emph{from scratch} for 1,000,000 steps on the reconstructed mel-spectrograms produced by the trained VQ-VAE. This explicit conditioning on codec-induced artefacts improves robustness. We use the AdamW \cite{loshchilov2019adamw} optimizer with an initial learning rate of $2\times10^{-4}$, $\beta_1=0.9$, $\beta_2=0.95$, a weight decay of 0.01, and a batch size of 32. The learning rate schedule includes a 2,500-step linear warm-up followed by cosine decay \cite{loshchilov2017sgdr}. The total generator loss combines adversarial, feature matching, and mel-spectrogram reconstruction terms: $L_G = L_G^{\text{adv}} + \lambda_{\text{feat}}L_{\text{feat}} + \lambda_{\text{mel}}L_{\text{mel}}$, with weights $\lambda_{\text{feat}}=2$ and $\lambda_{\text{mel}}=45$, as in the original paper~\cite{kong2020hifigan}.

\section{Evaluation}

\begin{figure}[!ht]
\centering
\includegraphics[width=.9\textwidth]{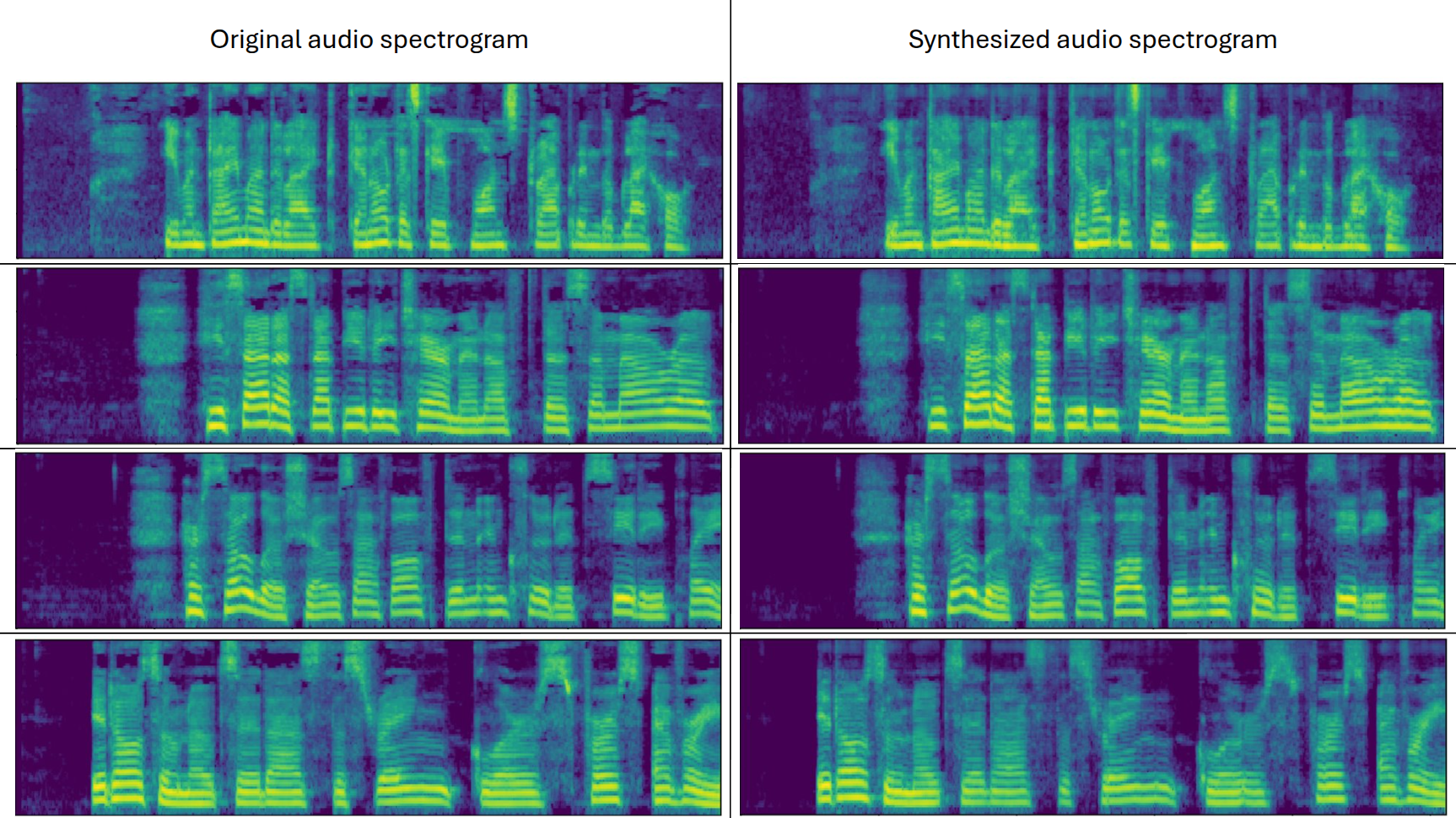}
\caption{Visual comparison of original mel spectrograms (left column) and their corresponding reconstructions from audio synthesized by our proposed codec (right column). Examples are randomly selected from the test set.}
\label{fig:spectograms}
\end{figure}

\subsection{Test set and metrics}
Evaluations were performed on 1,000 randomly drawn utterances from our multilingual speech corpus, comprising diverse speakers, genres, and utterance lengths. Each utterance was resampled to the appropriate input rate of the codec under test. We report five objective metrics:
\begin{itemize}
    \item \textbf{Mel‑Cepstral Distance (MCD):} A perceptually motivated measure comparing mel‑cepstral coefficients between reference and degraded speech~\cite{kubichek1993mcd}.
    \item \textbf{STOI:} The short‑time objective intelligibility measure, which correlates highly with speech intelligibility~\cite{taal2011stoi}.
    \item \textbf{PESQ:} Perceptual evaluation of speech quality, modeling human perception across various network conditions~\cite{rix2001pesq}. We use reference-less models from TorchAudio‑SQUIM~\cite{kumar2023squim}.
    \item \textbf{ViSQOL:} A full‑reference speech and audio quality metric available as open source, positioned as a lightweight alternative to PESQ/POLQA~\cite{chinen2020visqol}.
    \item \textbf{Real‑Time Factor (RTF):} The ratio between processing time and signal duration, measured on an NVIDIA RTX~4090 GPU.
\end{itemize}

All baselines were run using publicly available checkpoints. DAC models at 8~kbits/s (24~kHz and 16~kHz), EnCodec models at 12~kbits/s and 6~kbits/s, and the SNAC model at 0.98~kbits/s were obtained from their open-source repositories~\cite{kumar2023dac,defossez2022encodec,siuzdak2024snac}. Our codec operates at approximately 7.5~kbits/s with a sampling rate of 16~kHz.

\subsection{Objective comparison}
Table~\ref{tab:metrics} summarizes the average metrics over the 1,000‑utterance test set. Lower MCD and higher scores for STOI, PESQ and ViSQOL indicate better perceptual quality; lower RTF indicates faster inference. Values in parentheses denote the codec’s sampling rate and nominal bitrate.

\begin{table}[t]
  \centering
  \caption{Objective evaluation of neural codecs on our 1,000‑utterance test set. MCD is the mel‑cepstral distance (lower is better), PESQ and STOI measure perceptual quality and intelligibility (higher is better), ViSQOL is a full‑reference audio quality metric (higher is better) and RTF denotes the real‑time factor (lower is better). The ``Rate'' column lists the sampling rate in kHz and the nominal bitrate in kbits/s for each model.}
  \label{tab:metrics}
  \begin{tabular}{@{}lcccccc@{}}
    \toprule
    Model & Rate (kHz/(kbits/s)) & MCD $\downarrow$ & PESQ $\uparrow$ & RTF $\downarrow$ & STOI $\uparrow$ & ViSQOL $\uparrow$ \\
    \midrule
    DAC & 24/8 & \textbf{19.89} & 2.83 & 0.019 & \textbf{0.965} & \textbf{4.39} \\
    DAC & 16/8 & 40.99 & \textbf{2.92} & 0.017 & 0.874 & 4.01 \\
    EnCodec & 22/12 & 99.61 & 2.54 & 0.0066 & 0.954 & 3.26 \\
    EnCodec & 22/6 & 110.84 & 2.24 & 0.0088 & 0.921 & 2.90 \\
    SNAC & 24/0.98 & 91.23 & 2.86 & \textbf{0.0046} & 0.821 & 2.22 \\
    \textbf{Ours} & 16/7.5 & 133.53 & 2.70 & 0.013 & 0.844 & 2.82 \\
    \bottomrule
  \end{tabular}
\end{table}

\subsection{Discussion}
The results in Table~\ref{tab:metrics} show that our proposed codec, while not outperforming state-of-the-art models like DAC on every metric, achieves a competitive trade-off between quality, latency, and architectural simplicity. Its performance is comparable to EnCodec at 6 kbits/s, particularly in terms of PESQ and STOI, demonstrating strong perceptual quality and intelligibility.

The key advantage of our model is its single-stage, non-residual quantization scheme, which simplifies the architecture and may offer benefits in certain deployment scenarios. The objective results confirm that this simpler approach can still yield high-quality speech. While models like DAC achieve superior performance, they rely on more complex RVQ stacks. Therefore, our model should not be seen as a direct replacement for the highest-fidelity codecs, but rather as an effective and open foundation for further research. It provides a strong baseline for exploring improvements in single-stage quantization and vocoder-codec co-design.

To complement the objective evaluation, Figure \ref{fig:spectograms} provides a visual comparison between original mel spectrograms and those generated by our full codec pipeline. These examples, drawn from the held-out test set, demonstrate that our proposed architecture preserves the fundamental spectral structure with high fidelity, including the formant contours and the overall temporal envelope. This strong visual correspondence is indicative of the generated audio's intelligibility. A closer inspection reveals a slight smoothing effect in the higher frequencies, where fine-grained harmonic details are attenuated—a common artifact in adversarial synthesis systems. This visual evidence aligns well with the objective metrics in Table \ref{tab:metrics}, supporting the competitive STOI and PESQ scores and confirming the high quality of the synthesis.

\subsection{Bitrate derivation for our configuration}
\label{sec:bitrate}
With 16~KHz sampled audio and a hop of 128 samples, there are $16,000/128=125$ mel frames per second. Temporal downsampling by four yields $125/4=31.25$ latent steps per second. With $F{=}80$ mel bands and pooling by four along frequency, the latent grid has $F/4=20$ bands. Thus, the number of tokens per second is $31.25\times20=625$. Using a single codebook with $K{=}4096$ entries ($\log_2 K=12$ bits per token), the bitrate is $625\times12=7,500$~bps $\approx$\,\textbf{7.5~kbits/s}. This derivation guides our choice of baselines with bitrates between 6 and 9~kbits/s.

\section{Real‑Time Factor}
Real‑time factor (RTF) is defined as the ratio of wall‑clock processing time to signal duration. Table~\ref{tab:metrics} summarises the measured RTFs for each model. All codecs achieve $\mathrm{RTF}<0.03$, indicating real‑time processing on an RTX~4090 GPU. Our codec’s RTF of 0.013 arises from the moderate complexity of the HiFi‑GAN vocoder trained on reconstructed spectrograms.

\section{Conclusion and Future Work}
We introduced a low‑latency speech codec that leverages \textbf{single‑level, 2D block quantization} over mel patches and a HiFi‑GAN vocoder trained from scratch on reconstructed spectrograms. Our system attains a practical bitrate of $\sim$7.5~kbits/s with intelligibility and perceptual quality competitive with residual‑quantized codecs such as EnCodec and SNAC, while maintaining real‑time synthesis. The evaluation on a diverse 1,000‑utterance test set reveals the strengths and trade‑offs of several neural audio codecs when measured with STOI, MCD, PESQ, ViSQOL and RTF.

\paragraph{Future work.}
Our results open several avenues for future research. One key direction is to explore methods for further compressing the acoustic tokens. This could involve reducing the dimensionality of the frequency axis in the latent grid, potentially down to a 1D sequence of tokens, which would significantly lower the bitrate. Another promising area is to improve the VQ-VAE encoder's quality to produce a more robust representation, thereby simplifying the task for the subsequent vocoder and potentially improving overall fidelity. We also plan to investigate end-to-end joint training of the codec and vocoder, as well as exploring autoregressive models that operate with the encodings produced by our model so we can assess other aspects of our work.


\appendix
\section{Bitrate calculation}
\label{app:bitrate}

The bitrate of our codec can be computed analytically. Let the sampling rate be $\mathit{sr}=16,000$~Hz and the hop size be $h=128$ samples. Each second of audio therefore contains $\mathit{sr}/h=125$ mel frames. Applying a temporal downsampling factor of 4 reduces the latent frame rate to $125/4=31.25$ frames per second. Along the frequency axis we use $F=80$ mel bands and downsample by a factor of 4, yielding $20$ frequency bands. Thus the number of tokens produced per second is

\[
\text{tokens per second} = \frac{\mathit{sr}}{h\,\times\,\text{downsample time}} \times \frac{F}{\text{downsample freq}} = 31.25 \times 20 = 625.
\]

Each token encodes one of $K=4096$ codebook entries, which requires $\log_2 K = 12$ bits. The bitrate in bits per second is therefore

\[
\text{bitrate} = \text{tokens per second} \times \log_2 K = 625 \times 12 = 7500~\text{bps} \approx 7.5~\text{kbits/s}.
\]

This analytic derivation is consistent with the derivation presented in Sec.~\ref{sec:bitrate} of the main paper.

\section{Architecture details}
\label{app:architecture}

\begin{figure}[!ht]
\centering
\includegraphics[width=.9\textwidth]{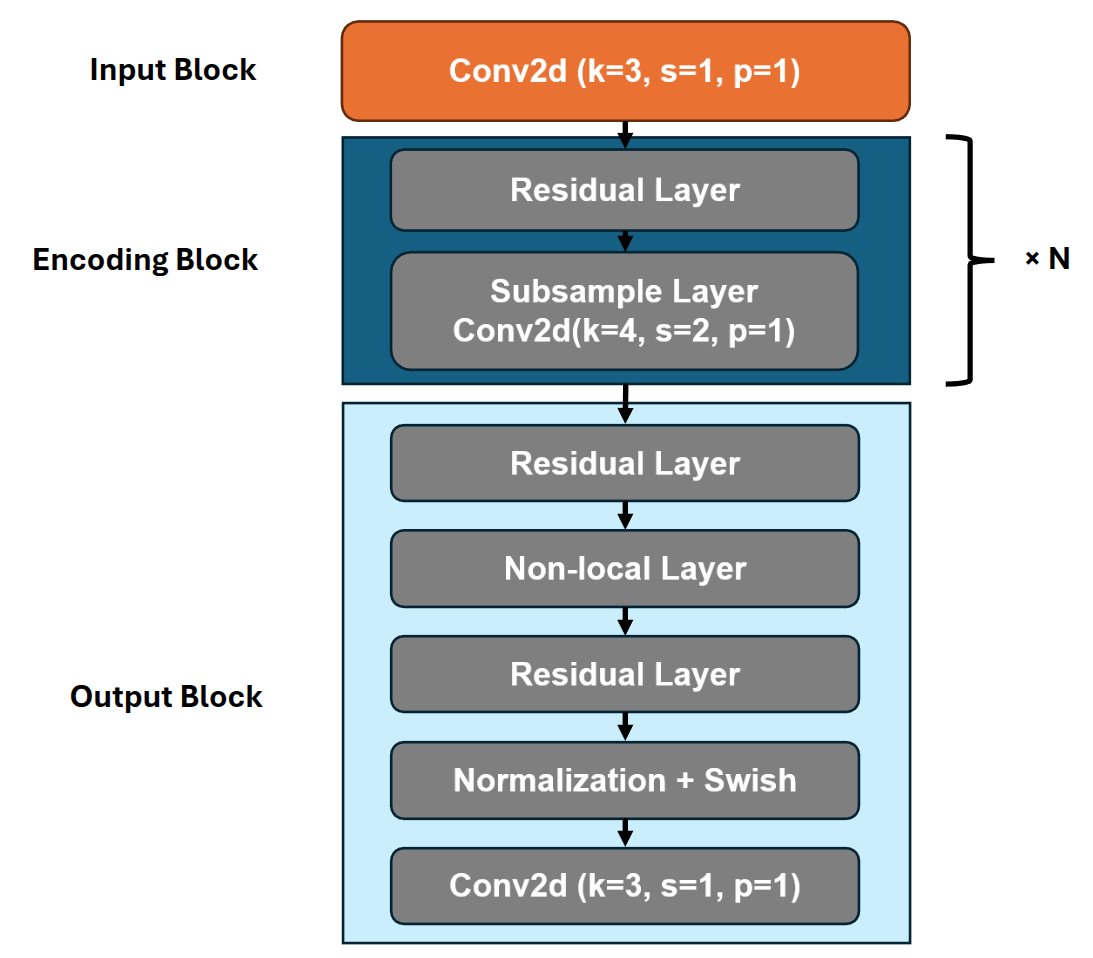}
\caption{Details of the blocks used inside de VA-VAE encoder. The same configuration is used by the decoder.}
\label{fig:visao_autoencoder}
\end{figure}

\begin{figure}[!ht]
\centering
\includegraphics[width=.49\textwidth]{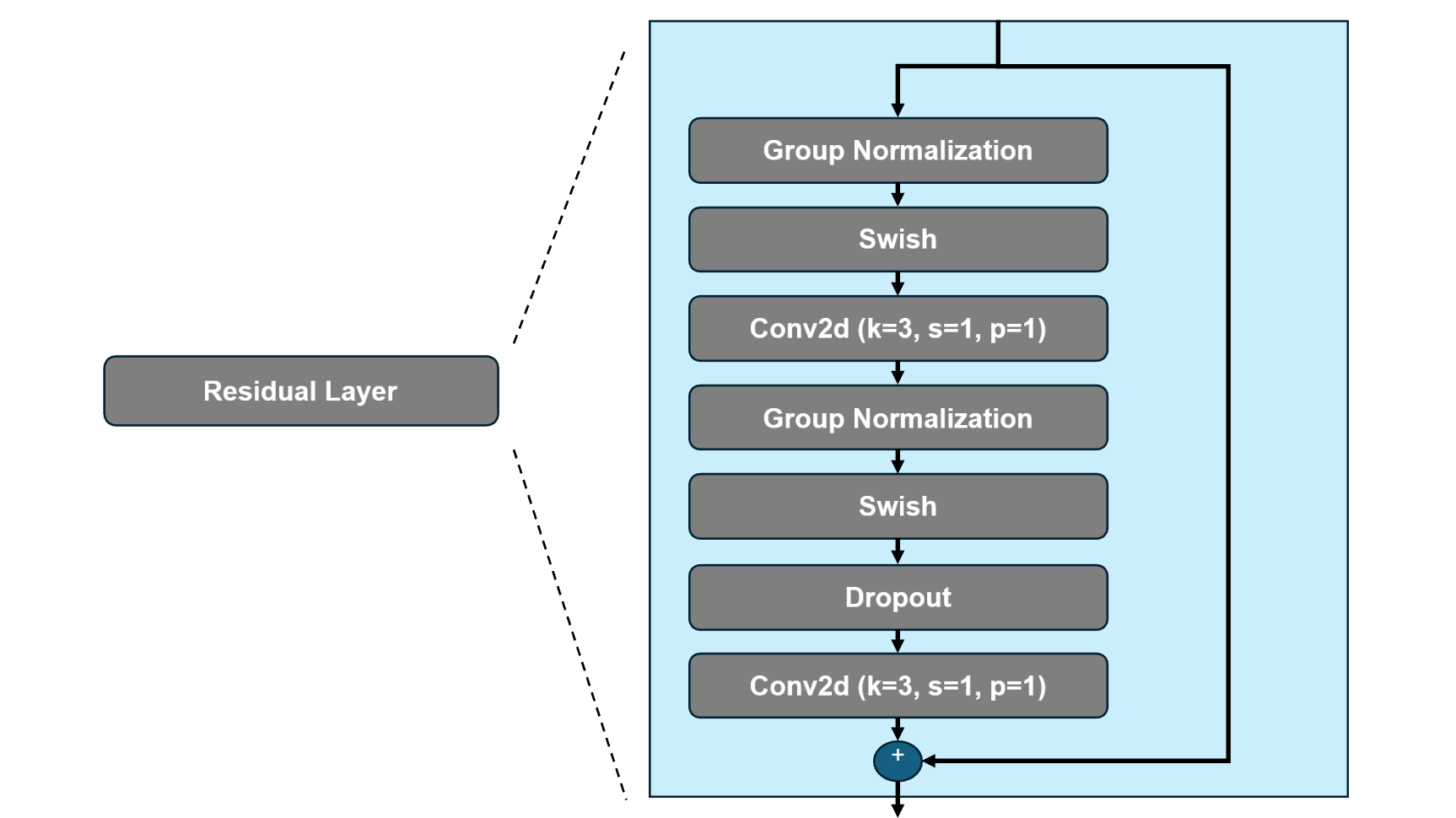}
\includegraphics[width=.49\textwidth]{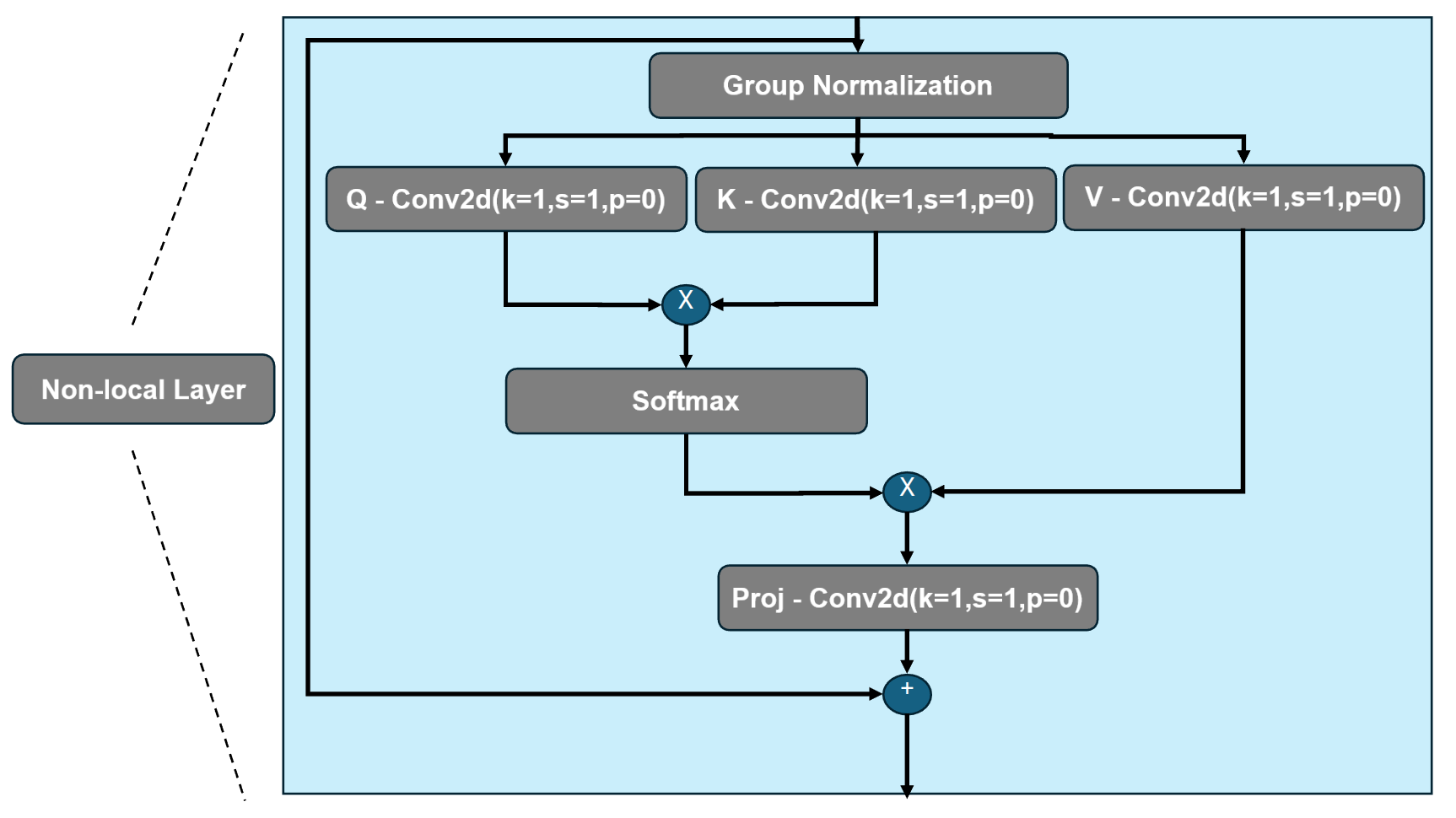}
\caption{Detailed view of the residual (left) and non-local (right) layers used within the VQ-VAE blocks.}
\label{fig:detalhe_autoencoder}
\end{figure}

\end{document}